\documentclass{elsart1p}

\usepackage{hyperref}
\usepackage[dvips]{graphicx}

\usepackage{bbm}
\usepackage{amsmath,amssymb,mathrsfs}
\usepackage[rflt]{floatflt}

\newcommand{\qs}{Q_\mathrm{s}}
\newcommand{\lqcd}{\Lambda_{_{\rm QCD}}}
\newcommand{\as}{{\alpha_\mathrm{s}}}

\newcommand{\xt}{{\boldsymbol{x}_\perp}}

\newcommand{\yt}{{\boldsymbol{y}_\perp}}

\newcommand{\ut}{{\boldsymbol{u}_\perp}}

\newcommand{\pt}{{\boldsymbol{p}_\perp}}
\newcommand{\qt}{{\boldsymbol{q}_\perp}}
\newcommand{\kt}{{\boldsymbol{k}_\perp}}

\newcommand{\nabt}{\boldsymbol{\nabla}_\perp}

\newcommand{\nr}[1]{(\ref{#1})} 
\newcommand{\ud}{\mathrm{d}}

\newcommand{\eq}{Eq.~}

\def\p{{\boldsymbol p}}
\def\x{{\boldsymbol x}}
\def\y{{\boldsymbol y}}
\def\u{{\boldsymbol u}}
\def\v{{\boldsymbol v}}
\def\k{{\boldsymbol k}}

\begin{document}

\begin{frontmatter}

\title{The glasma initial state and JIMWLK factorization}
\date{9 October 2008}

\author[cern,cea]{F. Gelis}
\author[cea]{T. Lappi}
\author[bnl]{R. Venugopalan}
\address[cern]{Theory Division, PH-TH, Case C01600, CERN,
 CH-1211, Geneva 23, Switzerland}
\address[cea]{Institut de Physique Th\'eorique,
  CEA/DSM/Saclay, B\^at. 774,
  91191, Gif-sur-Yvette Cedex, France}
\address[bnl]{Physics Department, Brookhaven National Laboratory
  Upton, NY-11973, USA}

\begin{abstract}
We review recent work on understanding the next to leading order
corrections to the classical fields that dominate the initial 
stages of a heavy ion collision. We have recently shown that the
leading $\ln 1/x$ divergences of these corrections to gluon multiplicities
can be factorized into the JIMWLK evolution of the color charge density
distributions.
\end{abstract}

\end{frontmatter}

\section{Introduction: Glass and Glasma}

At large energies (small $x$)
the hadron or nucleus wavefunction is characterized by a saturation scale 
$\qs \gg \lqcd$ arising from the strong nonlinear interactions of the color field.
In the Color Glass Condensate (CGC) (for reviews see \cite{Iancu:2003xm,Weigert:2005us})
 framework the small $x$ part of the hadron wavefunction is described in terms
of a classical Weizs\"acker-Williams (WW) field radiated by the
hard, large $x$, sources.
The color sources $\rho$ are stochastic variables fluctuating according to a probability 
distribution $W_y[\rho(\xt)]$, where $y$ is the rapidity scale 
separating fast and slow partons~\cite{McLerran:1994ni}.

 The matter during the first fraction of a fermi
in a collision of two such objects is what we refer to as the Glasma \cite{Lappi:2006fp}.
The glasma configuration after the collision, at times $0 \leq \tau \lesssim 1/\qs$,
consists of longitudinal chromomagnetic and -electric field which depend on the transverse 
coordinate on a typical scale $\sim 1/\qs$. 
As the system expands the fields are diluted and can 
be treated as particles, forming the leading order (LO) production is the contribution
that is computed in the numerically solving the classical Yang-Mills 
equations~\cite{Krasnitz:1998ns,Krasnitz:2001qu,Lappi:2003bi}.

In the following we are concerned with the next to leading
order (NLO) in
$g$, $\hbar$ or, equivalently, loop corrections to this classical field. At NLO one can produce pairs
of quarks (see Refs.~\cite{Gelis:2003vh,Gelis:2004jp,Gelis:2005pb}) or gluons (real corrections)
and one must take into account one loop corrections to  the 
classical field (virtual corrections). We shall argue that these
corrections have logarithmically divergent
contributions, which must then be resummed into the renormalization group evolution of
the sources $W[\rho(\xt)]$~\cite{Gelis:2008rw,Gelis:2008ad}.

\section{Factorization theorem}
\label{sec:factgeneral}

It is perhaps useful to look first at the weak field
limit of the CGC, where particle production can be computed using  $k_T$-factorization
(\cite{Gribov:1984tu}, see e.g.~\cite{Kharzeev:2003wz}
for an application to heavy ion collisions).
The leading order multiplicity  is
\begin{equation}
\frac{\ud N}{\ud^2\pt \ud y} =
\frac{1}{\as}\frac{1}{\pt^{\!\!\!2}} \int \frac{\ud^2\kt_1}{(2 \pi)^2}
\varphi_y(\kt_1) \varphi_y(\kt_2) \delta^2(\kt_1 + \kt_2 - \pt).
\end{equation}
For the real part of the Leading Log correction to this result one must take the
corresponding expression for double inclusive gluon production
\begin{equation}
\frac{\ud N}{\ud^2\pt \ud y_p \ud^2\qt \ud y_q} =
\frac{1}{\as}\frac{1}{\pt^{\!\!\!2} \qt^{\!\!\!2}} \int \frac{\ud^2\kt_1}{(2 \pi)^2}
\varphi_y(\kt_1) \varphi_y(\kt_2) \delta^2(\kt_1 + \kt_2 - \pt - \qt).
\end{equation}
and integrate it over the phase space of the second gluon $(\qt,y_q)$. 
Note that at leading log accuracy we have here taken
the multi-Regge kinematical limit, assuming
that the two produced gluons are far apart in rapidity (see 
e.g.~\cite{Leonidov:1999nc}).
The integral over $y_q$ diverges linearly (this is the general behavior of the $gg \to gg$ scattering 
amplitude in the high energy limit $t$ fixed, $s\sim -u \to \infty$). 
This divergence is compensated (to the appropriate order in $\as$) by the real part of 
the BFKL evolution equation for $\varphi_y(\kt_1)$.

\begin{floatingfigure}{0.3\textwidth}
\includegraphics[width=0.3\textwidth]{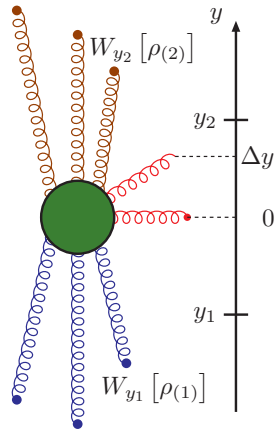}
\caption{Production of two gluons: the integral over $\Delta y$ is divergent.}
\label{fig:facth}
\end{floatingfigure}

In the fully nonlinear case of AA collisions the $k_T$-factorization is broken
(see e.g.~\cite{Krasnitz:1998ns,Blaizot:2008yb}), and
one must solve the equations of motion to all orders in the strong classical field.
The analogue of the unintegrated parton distributions $\varphi_y(\kt)$ is
the color charge density distribution $W_y[\rho]$. These are similar in the
sense that they are not (complex) wavefunctions but (at least loosely speaking)
real probability distributions. Factorization can be understood as a statement that 
one has found a convenient set of degrees of freedom 
in which one can compute physical observable from only the diagonal
elements of the density matrix of the incoming nuclei.
The difference is that when in the dilute case these degrees of freedom are
 numbers of  gluons with a given momentum, in the nonlinear case the
appropriate variable  is the color charge
density and the relevant evolution equation is JIMWLK, not BFKL.
The kinematical situation, however, remains the same. To produce a gluon at a very large 
rapidity (or a contribution in the loop integral of the virtual contribution with a large
$k^+$) one must get a large $+$-momentum from the right-moving source. 
Thus one is probing the  source at a large $k^+$, i.e. small distances in
$x^-$, and the result must involve $W_y[\rho]$ at a larger rapidity.
The underlying physical interpretation of factorization is that this
fluctuation with a large $k^+$ requires such a long interval in $x^+$ 
to radiated that it must be produced well before and independently of the
interaction with the other 
(left moving and thus localized in $x^+$) source.
 The concrete task is then to show that when one computes the NLO corrections
to a given observable in the Glasma, all the leading 
logarithmic divergences can be absorbed into the RG evolution of the sources
with the same Hamiltonian that was derived by considering only the DIS process. 
This is the  proof~\cite{Gelis:2008rw,Gelis:2008ad,Gelis:2007kn} of factorization that 
we will briefly describe in the following.

\section{Deriving JIMWLK factorization}
\label{sec:proof}

Consider the single inclusive gluon multiplicity
which is a sum of probabilities to produce $n+1$ particles, with the phase space
of the additional $n$ must be integrated out
\begin{equation}
\frac{d N}{d^3\vec\p}\sim \sum_{n=0}^\infty 
\frac{1}{n!}\int\Big[d^3\vec\p_1\cdots d^3\vec\p_n\Big]\;
\left|\big<{\vec\p}\;\,\vec\p_1\cdots\vec\p_n\big|0\big>\right|^2.
\end{equation}

Because we have a theory with external color sources of order $\rho \sim 1/g,$
all insertions of source appear at the same order~\cite{Gelis:2006yv}. 
A calculation using the 
Schwinger-Keldysh formalism leads to the following results:
At LO, the multiplicity is obtained from the  \emph{retarded} solution of classical 
field equations (here $(\dots)$ includes the appropriate normalization and projection
to physical polarizations)
\begin{equation}
\frac{\ud N_{_{\rm LO}}}{d^3\vec\p}
=
\!
\int \! \ud^3 \x \ud^3 \y
\;e^{i\vec\p\cdot(\vec\x-\vec\y)}\;
(\cdots)
\Big[
{{\cal A}^\mu(t,\vec\x)}{{\cal A}^\nu(t,\vec\y)}
\Big]
\Big|_{t\to\infty}.
\end{equation}
The NLO contribution includes the one loop correction to the classical field and the 
$+-$ component of the Schwinger-Keldysh (SK) propagator in the background field
\begin{equation}
\frac{\ud N_{_{\rm NLO}}}{d^3\vec\p}
\!\!
=
\!
\int_{\x,\y} \! \!\! 
\;e^{i\vec\p\cdot(\vec\x-\vec\y)}\;
(\cdots)
\Big[
{\mathcal{G}_{+-}^{\mu\nu}(x,y)}
+
{\beta_+^\mu(t,\vec\x)}\;{\mathcal{A}_-^\nu(t,\vec\y)}
+
{\mathcal{A}_+^\mu(t,\vec\x)}\;{\beta_-^\nu(t,\vec\y)}
\Big]
\Big|_{t\to\infty}.
\end{equation}
\begin{floatingfigure}{0.45\textwidth}
\begin{center}
\includegraphics[width=0.22\textwidth]{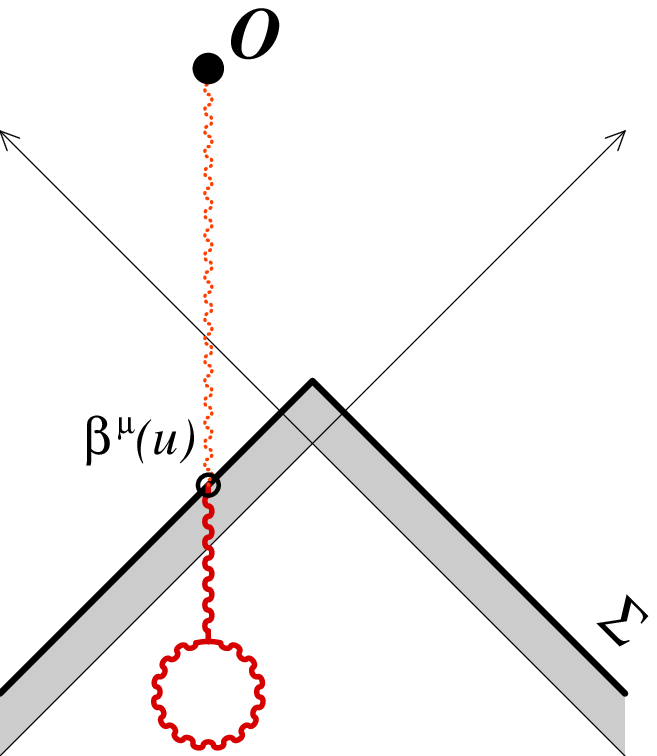}
\includegraphics[width=0.22\textwidth]{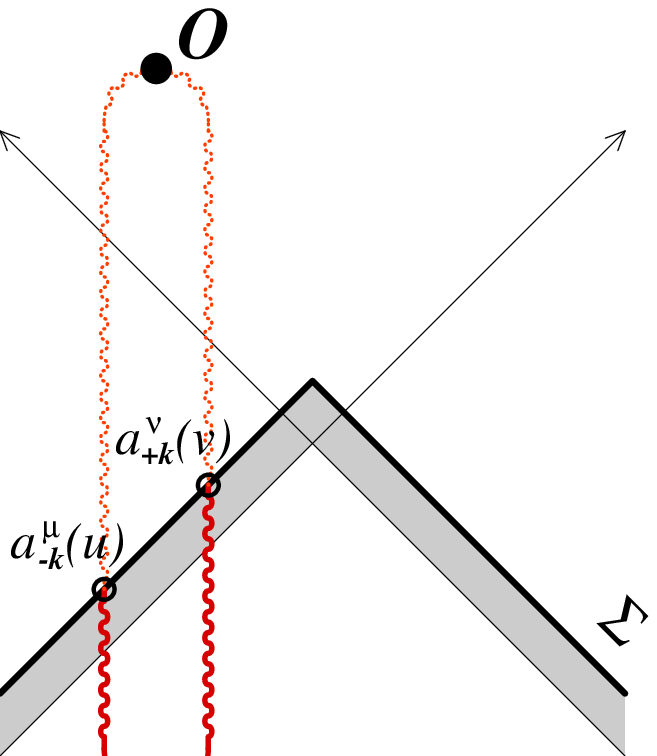}
\end{center}
\caption{The 1-loop one and two point functions in the background field.}
\label{fig:betagpm}
\end{floatingfigure}
Because of the SK index structure (also $\beta$ satisfies an equation of motion with
 a retarded  boundary condition), one can express the propagation of a small
fluctuation $a^\mu(x)$ above the  past light cone $\Sigma$ as 
a functional derivative $\mathbb{T}_{\u}$ of the LO classical field $\mathcal{A}^\mu(x)$ 
with respect to its  initial condition on $\Sigma$:
$a^\mu(x) =\int_{\vec\u\in{\rm \Sigma}} a(\vec\u) \cdot
 \mathbb{T}_{\u} \mathcal{A}^\mu(x).$
This leads after some rearrangements to the expression for the NLO contribution to the multiplicity
as a functional derivative operator acting on the leading order result:
\begin{equation} \label{eq:dop}
\left.\frac{dN}{d^3\vec\p}\right|_{_{\rm NLO}}
=
\bigg[
\frac{1}{2} \! \int_\Sigma \! \ud^3\u \ud^3\v
\mathcal{G}_{\mu \nu}(\u,\v) \mathbb{T}^\mu_{\u} \mathbb{T}^\nu_{\v}
+\int_\Sigma \! \ud^3\u
\mathbf{\beta}_\mu(\u) \mathbb{T}^\nu_{\u}
\bigg]\;
\left.\frac{dN}{d^3\vec\p}\right|_{_{\rm LO}}. 
\end{equation}
Here the two point function below the light cone
$
\mathcal{G}^{\mu\nu}(\vec\u,\vec\v)\equiv
\int\frac{d^3\vec\k}{(2\pi)^3 2E_\k}\; a^\mu_{-\k}(\u)\,a^\nu_{+\k}(\v)
$
is bilinear in the small fluctuation field $a^\mu(x)$ satisfying the small
fluctuation equation of motion with and initial condition given by a plane wave
$\lim_{x^0\to -\infty}a^\mu_{\pm \k}(x) = \epsilon^\mu(\k)e^{\pm i k\cdot x}$;
see Fig.~\ref{fig:betagpm} for a pictorial representation of this structure.

The leading logarithmic contribution comes from the longitudinal component of the 
integral over $\k$, the momentum of the initial plane wave perturbation (and the 
corresponding momentum in the one loop source term for the equation of motion
satisfied by $\beta$). This LLog part of the functional derivative \nr{eq:dop} operator
turns out to be precisely equivalent to the sum of the JIMWLK Hamiltonians
describing the RG evolution of the source distributions $W_y[\rho]$.
The fact that no other terms with the same logarithmic divergences appear is the 
crucial result for factorization. The JIMWLK  Hamiltonian 
\begin{equation}\label{eq:hjimwlk}
\mathcal{H}
\equiv \frac{1}{2}
\int \! \ud^2\xt \ud^2\yt 
D_a(\xt)
\eta^{ab}(\xt,\yt)
D_b(\yt)
\end{equation}
is most naturally expressed in terms of Lie derivatives $D_a(\xt)$ operating on
 the Wilson lines formed from the classical field
($ U(\xt) = \mathrm{P} \exp \left\{-i g \int \ud x^-  \left(1/\nabt^2 \right) \rho(\xt,x^-) \right\} $
for the nucleus moving in the $+z$ direction)
in terms of which the kernel in \eq\nr{eq:hjimwlk} is
\begin{equation}
\eta^{ab}
(\xt,\yt)
=
\frac{1}{\pi}
\int_{\ut}
\!\!
\frac{(\xt-\ut)\cdot(\yt-\ut)}{(\xt-\ut)^2(\yt-\ut)^2}\;
\Big[
U_\xt U^\dag_\yt
-U_\xt U^\dagger_\ut
-U_\ut U^\dag_\yt
+1\Big]^{ab} 
\label{eq:eta-f}.
\end{equation}

Let us conclude by summarizing some important aspects of the JIMWLK factorization theorem 
of Refs.~\cite{Gelis:2008rw,Gelis:2008ad}. We are interested in the high energy kinematical 
limit, with transverse momenta $\sim \qs$ and the energy $s$  large; 
the dilute limit is that of BFKL physics. The relevant degrees of freedom in
this framework are the color charge densities $\rho$ of the fast partons and the classical fields
of the small $x$ ones. The color charges $\rho \sim 1/g$ are parametrically large, 
and thus
the problem is inherently nonperturbative, but there is,
however, a consistent weak coupling 
or loop expansion.
The primary observables of interest are inclusive single and multigluon multiplicities 
(not cross sections for producing a fixed number of particles), which leads to results that 
can be expressed in terms of retarded (and advanced) propagators. We express these propagators
as functional derivatives with respect to the field on an initial surface on the light cone,
which can then be mapped to the functional derivatives of the color fields of the
individual nuclei appearing in the JIMWLK Hamiltonian.

\emph{Acknowledgments:} R. V.'s work is supported
by the US Department of Energy under DOE Contract No.
DE-AC02-98CH10886. F.G.'s work is supported in part by Agence
Nationale de la Recherche via the programme ANR-06-BLAN-0285-01.

\bibliographystyle{h-physrev4mod2}
\bibliography{spires}
\end{document}